\documentclass[pra,twocolumn,showpacs,lengthcheck,preprintnumbers,longbibliography,superscriptaddress]{revtex4-2}
\usepackage{epsfig}
\usepackage{amsmath}
\usepackage{amsfonts}
\usepackage{fancyhdr}
\usepackage{amsmath}
\usepackage{amssymb}
\usepackage{mathtools}
\usepackage{graphicx}
\usepackage{siunitx}
\usepackage[unicode]{hyperref}
\hypersetup{
   unicode=true,          
   plainpages=false,
   colorlinks=true,       
   citecolor=blue,        
}
\begin{document}

\title{Stability analysis and attractor dynamics of 3D dark solitons with localized dissipation}
\author{Christian Baals}
\affiliation{Department of Physics and Research Center OPTIMAS, Technische 
Universität Kaiserslautern, 67663 Kaiserslautern, Germany}
\affiliation{Graduate School Materials Science in Mainz, Staudinger Weg 9, 55128 Mainz, Germany}
\author{Alexandre Gil Moreno}
\affiliation{Department of Physics and Research Center OPTIMAS, Technische 
Universität Kaiserslautern, 67663 Kaiserslautern, Germany}
\author{Jian Jiang}
\affiliation{Department of Physics and Research Center OPTIMAS, Technische 
Universität Kaiserslautern, 67663 Kaiserslautern, Germany}
\author{Jens Benary}
\affiliation{Department of Physics and Research Center OPTIMAS, Technische 
Universität Kaiserslautern, 67663 Kaiserslautern, Germany}
\author{Herwig Ott}
\affiliation{Department of Physics and Research Center OPTIMAS, Technische 
Universität Kaiserslautern, 67663 Kaiserslautern, Germany}

\date{\today}

\begin{abstract}
We study the stability and the attractor dynamics of an elongated Bose-Einstein condensate with dark or grey kink solitons in the presence of localized dissipation.
To this end, the 3D Gross-Pitaevskii equation with an additional imaginary potential is solved numerically.
We analyze the suppression of the snaking instability in dependence of the dissipation strength and extract the threshold value for the stabilization of the dark soliton for experimentally realistic parameters. 
Below the threshold value, we observe the decay into a solitonic vortex.
Above the stabilization threshold, we observe the attractor dynamics towards the dark soliton when initially starting from a grey soliton.
We find that for all initial conditions the dark soliton is the unique steady-state of the system - even when starting from the BEC ground state.
\end{abstract}

\maketitle

\section{Introduction}

Dissipative processes, such as losses or decoherence, are usually considered as a nuisance in quantum systems, because they tend to destroy the coherence and drive the system from quantum to classical behavior \cite{Breuer_book}.
This can become relevant even in well isolated quantum systems such as ultracold atomic gases, where dissipative processes are strongly suppressed, but never absent.
In recent years, open system control of quantum matter has emerged as a new field of research, which perceives dissipative processes as a resource for quantum engineering and state preparation \cite{Diehl2008, Diehl2010,PhysRevLett.110.035302,PhysRevLett.102.144101,CPA,Eur.Phys.J.D63.63-71, Witthaut2011,PhysRevA.84.041606,PhysRevLett.122.040402}.
Such an approach requires that the desired quantum state is the dark state/steady-state of the system's time evolution in the presence of an engineered dissipative process.
If the dark state/steady state is unique, the system will evolve towards it, independent of the initial condition.
This results in an attractor dynamics towards the steady-state.

Here, we perform a realistic numerical experiment to study the stabilization and attractor dynamics of a dark soliton in an elongated, three-dimensional atomic Bose-Einstein condensate.
We focus on two central questions:
\begin{enumerate}
\item Can the dissipation be engineered in such a way that the dark soliton is stabilized under the systems time evolution?
\item Does the dissipation induce attractor dynamics towards the steady-state, irrespective of the initial conditions?
\end{enumerate}

We will show in this paper, that both questions can be answered positively for experimentally realistic parameters.
Our study exemplifies the concept of open system control on a specific scenario and explicitly analyzes the emerging attractor dynamics.

The quantum system under consideration is a harmonically trapped Bose-Einstein condensate of atoms, which we describe in the mean-field limit by means of the 3D Gross-Pitaevskii equation (GPE).
We focus on dark kink-solitons (DS) \cite{PhysRevLett.83.5198,Becker2008,PhysRevA.101.053629,Frantzeskakis_2010} which are stationary solutions of the GPE but dynamically unstable for a large variety of trapping frequencies \cite{PhysRevA.60.R2665}.
Previous work in cylindrical trapping geometries has shown that the DS can decay into several different structures depending on the chemical potential and the radial trapping frequency \cite{Mateo_2015}.
Adding a local loss process, which we describe by an imaginary potential in the GPE, we study the time evolution of the DS for different strengths of the imaginary potential.
A similar situation has been studied for the 2D GPE.
Here it has been shown that adding a 1D Gaussian-shaped conservative repulsive potential can lead to the suppression of the snaking instability \cite{PhysRevA.82.023621}.
In one dimension, both, studies in a 1D GPE system \cite{PhysRevLett.102.144101} and in a Bose-Hubbard system \cite{Eur.Phys.J.D63.63-71} show the emergence of a stable DS under these conditions.
In addition, studies with $\mathcal{PT}$-symmetric dipoles in 1D GPE systems haye shown that for certain dissipation strengths moving light-grey solitons can be pinned \cite{Susanto_2015}.
Starting from different initial states and varying the applied dissipation strength, we map out the stability region for the DS, classify the emergent instability modes and characterize the attractor dynamics towards the DS.
This work is inspired by previous experiments in our group \citep{PhysRevLett.110.035302,Gericke2007}, which serve as a guideline for the chosen parameters.
Regarding the numerics, we efficiently solve the GPE on a GPU.

\section{System: 3D Gross-Pitaevskii equation with imaginary potential} \label{sec:sys}

We consider a BEC subject to local losses within a mean-field theory, where the condensate order parameter is described by the Gross-Pitaevskii equation with imaginary potential (IGPE)
\begin{equation}
i\hbar\frac{\partial \Psi}{\partial t} = \left( -\frac{\hbar^{2}}{2m}\nabla^{2} + V\left(\vec{r}\right) + g\left\vert\Psi\right\vert^{2} - \frac{i\hbar}{2}\gamma\left(\vec{r}\right)\right)\Psi,
\label{eq:IGPE}
\end{equation}
where $g=\frac{4\pi\hbar^{2}a}{m}$ is the interaction strength, $a$ is the $s$-wave scattering length, 
\begin{equation}
V\left(\vec{r}\right) = \frac{1}{2}m\left(\omega^{2}_{x}x^{2} + \omega^{2}_{y}y^{2} + \omega^{2}_{z}z^{2}\right)
\label{eq:3D_HO}
\end{equation}
is the 3D harmonic trapping potential and $\gamma\left(\vec{r}\right)$ describes the local particle losses.
The number of particles $N$ is given by the normalization condition $N = \int d\vec{r} \left\vert\Psi\right\vert^{2}$.

Such a scenario can be studied experimentally with a BEC and an additional scanning electron microscope \cite{PhysRevLett.110.035302, Gericke2007} where a tightly focused electron beam removes and ionizes atoms from the BEC, which are subsequently extracted by an electric field and detected.
This technique allows for a high resolution manipulation of the BEC with - apart from the losses - almost negligible back-action on the BEC for a relatively long time.
Previous studies \cite{PhysRevLett.110.035302,CPA} have shown that the IGPE [eq.(\ref{eq:IGPE})] is indeed an adequate model to describe the system.
To perform the numerical analysis with experimentally realistic parameters, we set the number of particles to $N=80\cdot 10^{3}$ and the trap frequencies to $\left(\omega_{x},\omega_{y},\omega_{z}\right) = 2\pi\cdot\left(\SI{12}{\Hz},\SI{170}{\Hz},\SI{170}{\Hz}\right)$.
We refer to the $x$-coordinate as the axial direction and to $y$ and $z$ as the radial directions.
We model the imaginary potential
\begin{equation}
\gamma\left(\vec{r}\right) = \gamma_{0}\exp\left(-\frac{\left(x-x_{0}\right)^{2}}{2 w^{2}_{\mathrm{diss}}}\right)
\label{eq:imagGauss}
\end{equation}
as a Gaussian profile along the $x$ axis (width $w_{\mathrm{diss}}$) and as constant along the $y$- and the $z$-axis.
Experimentally, this can be realized by scanning the electron beam (propagation direction $z$) along the $y$-direction much faster than any intrinsic timescale of the BEC.

We solve the IGPE [eq.(\ref{eq:IGPE})] numerically.
To this end, we implement a time-splitting spectral method \cite{BAO2002487} in the programming language Julia.
We use the package CUDA.jl \cite{besard2018juliagpu,besard2019prototyping} to run our simulations on a GPU (NVIDIA GeForce RTX 2060 Super) which results in a speed-up by a factor of about 20 compared to our CPU (HP Z620, 2x Intel XEON E5-2670).

\section{Stability of stationary kink solitons} \label{sec:stabilityDS}

Dark kink solitons are stationary but dynamically unstable solutions to eq. (\ref{eq:IGPE}) with $\gamma\left(\vec{r}\right)\equiv 0$.
They can be written as $\Psi\left(\vec{r},t\right) = \mathrm{e}^{-\frac{i}{\hbar}\mu t}\psi\left(\vec{r}\right)$ with the chemical potential $\mu$.
In order to find such a solution $\psi\left(\vec{r}\right)$ we follow the idea given in refs \cite{Mateo_2015,PhysRevLett.113.255302} and start with an ansatz 
\begin{equation}
\psi\left(\vec{r}\right) = \chi_{\mathrm{TF}}\left(\vec{r}\right)\tanh\left(\frac{x}{\xi\left(\vec{r}\right)}\right)
\label{eq:ansatzDS}
\end{equation}
in the Thomas-Fermi regime where $\chi_{\mathrm{TF}} = \sqrt{\frac{\mu_{\mathrm{loc}}}{g}}$ is the Thomas-Fermi wave function with the local chemical potential $\mu_{\mathrm{loc}} = \mu-V\left(\vec{r}\right)$ and the local healing length $\xi\left(\vec{r}\right) = \frac{\hbar}{\sqrt{m\mu_{\mathrm{loc}}\left(\vec{r}\right)}}$.
Starting from eq. (\ref{eq:ansatzDS}) we find the DS solution to eq. (\ref{eq:IGPE}) numerically employing imaginary time evolution.

\begin{figure}[t!]
\includegraphics[width=\linewidth]{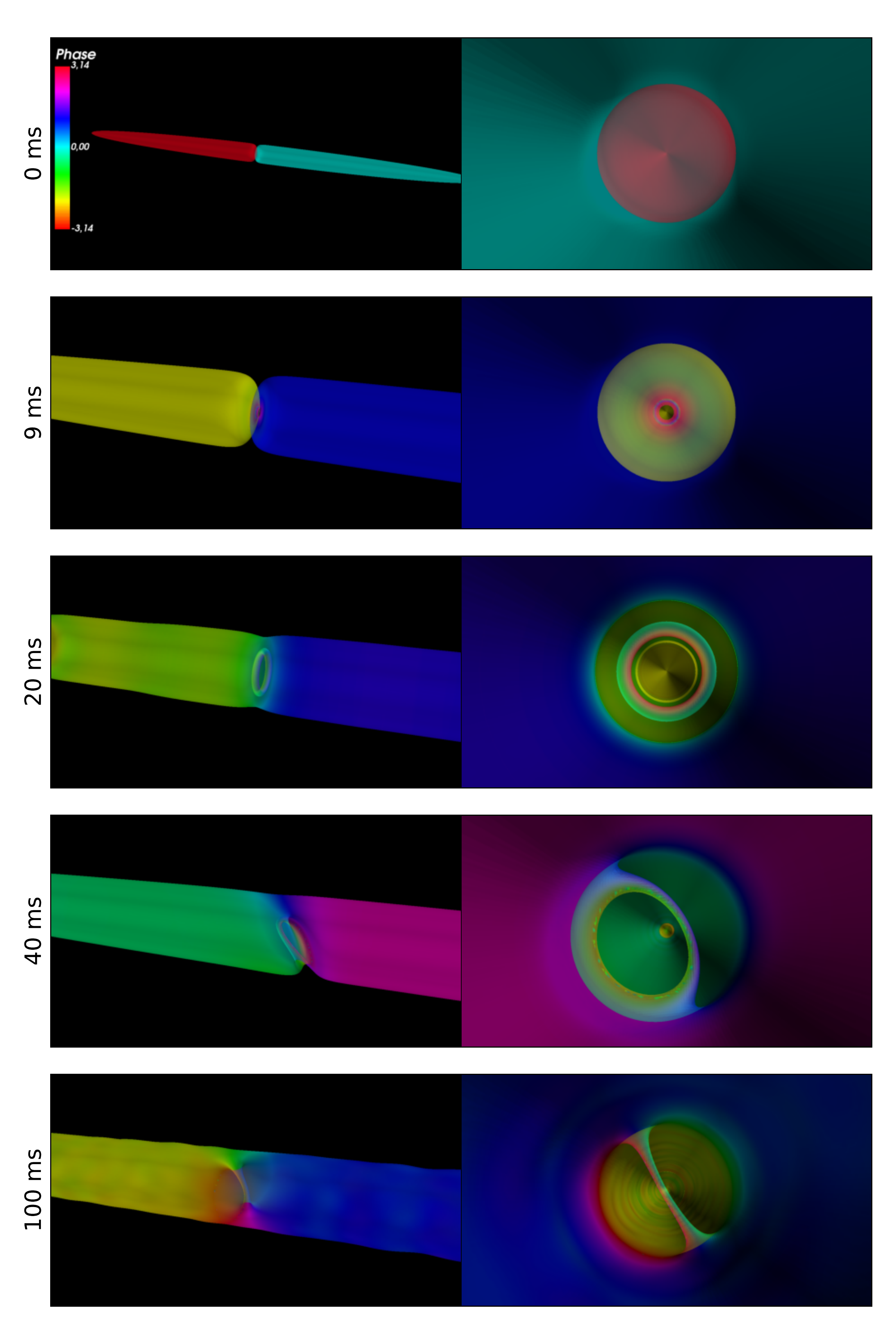}
\caption{Iso-surface plots of the DS without dissipation after different times.
We show the density at 10\% of its maximum as contours and the phase as color code. Top: initial DS at $t=0$\,ms.
After \SI{9}{\milli\second} the central plane of the DS starts to bend.
This is the beginning of the snaking instability.
After \SI{20}{\milli\second} the VR appears as an intermediate stationary state.
Being unstable itself (see the bending and shift after \SI{40}{\milli\second}), it eventually decays into a solitonic vortex (\SI{100}{\milli\second}), which is dynamically stable.
}
\label{fig:isoSurf_gamma=0}
\end{figure}

Having found the stationary DS, we first consider the time evolution without localized dissipation.
This has been studied already in refs \cite{Mateo_2015,PhysRevLett.113.255302} for a cylindrical trap.
For the elongated harmonic trap that we study here, we expect qualitatively similar results since the potential in the axial direction does not change on the length scale of the soliton.
Depending on the ratio of the chemical potential to the vibrational energy of the radial harmonic oscillator, the DS is either dynamically stable or unstable \cite{Mateo_2015}.
In our case we have $\mu=\hbar\cdot \SI{6600}{\Hz}$ and thus a ratio $\frac{\mu}{\hbar\omega_{\perp}} = 6.2$.
According to refs \cite{Mateo_2015,PhysRevLett.113.255302} we expect the DS to be dynamically unstable with two energetically lower lying states: a dynamically unstable single vortex ring (VR) and a dynamically stable solitonic vortex (SV).
Indeed, this is what we observe when evolving the DS in time.
Iso-surface plots of the density at characteristic points in time are shown in fig. \ref{fig:isoSurf_gamma=0}.

As a next step, we switch on the dissipation.
We here restrict ourselves to experimentally accessible values and choose $w_{\mathrm{diss}} = \SI{130}{\nano\meter}$ and $\gamma_{0}\in[0,7100]\,\si{\Hz}$ \cite{PhysRevLett.110.035302}.
\begin{figure}[t!]
\includegraphics[width=\linewidth]{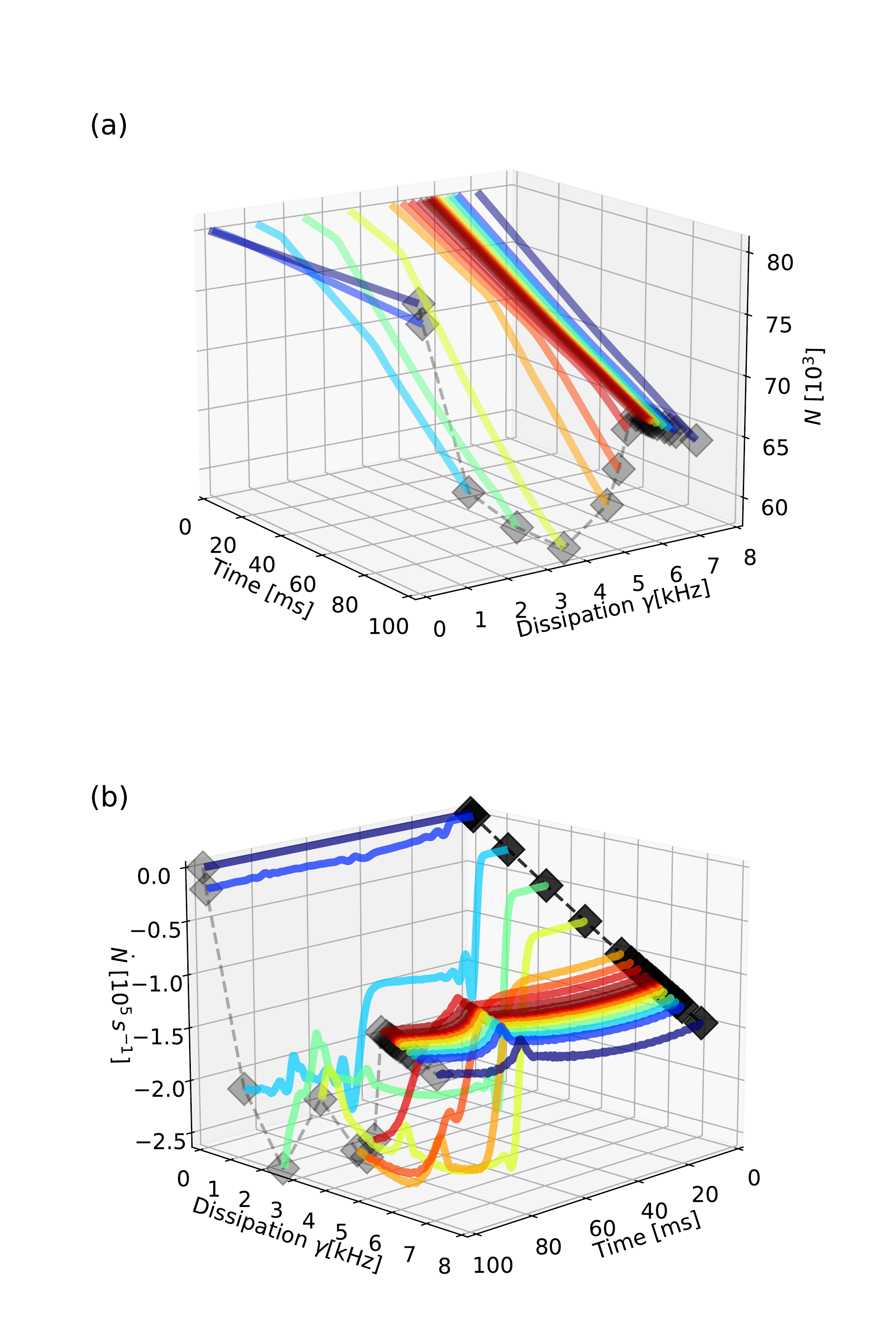}
\caption{Number of particles $N$ (a) and loss rate $\dot{N}$ (b) over time and dissipation strength $\gamma_{0}$.
For better visibility, we plot in (b) the loss rate as negative values and have exchanged the time and dissipation axis as compared to (a).
The local maximum of $\dot{N}$ at \SI{83}{\milli\second} is due to a density wave that is created by the dissipative potential.
It travels towards the edge of the BEC gets reflected there and returns towards the imaginary potential.}
\label{fig:N_dNdt_gamma_t}
\end{figure}
The amount of particles removed from the system during the time evolution is given by the overlap between the region of losses with the atomic density:

\begin{equation}
\dot{N}=-\int  \gamma(\vec{r})\left\vert\psi\left(\vec{r}\right)\right\vert^2 d^3 r.
\end{equation}

We can then identify the different stationary states (DS, VR and SV) from the loss rate of paricles, as each of them has a characteristic density overlap with the loss region.
In fig. \ref{fig:N_dNdt_gamma_t} we show the number of particles $N(t)$ in the condensate and the loss rate $\dot{N}(t)$ up to $t=100$\,ms for different dissipation strength $\gamma_{0}$.
Looking at $N(t)$ we see phases of linear decrease, connected by kinks.
This behavior becomes more obvious for $\dot{N}(t)$, where this translates into plateaus.
Up to $\gamma_{0}=\SI{2.4}{\kilo\Hz}$ two plateaus in $\dot{N}$ (or two kinks in $N$) are visible.
The first plateau with the lowest loss rate corresponds to the DS which has the lowest density overlap with the imaginary potential.
The second plateau corresponds to the VR, which shows a higher loss rate.
The highest loss rate is observed for the SV towards the end of the simulation time.
Between $\gamma_{0}=\SI{2.4}{\kilo\Hz}$ and $\gamma_{0}=\SI{5.2}{\kilo\Hz}$ we observe only the decay into a VR within our computational time of \SI{100}{\milli\second}.
For $\gamma_{0}>\SI{5.2}{\kilo\Hz}$ no decay is visible, which indicates the onset of stabilization of the DS.
This can also be seen in the behaivour of $N(t)$, where the remaining number of atoms after $t=100$\,ms first decreases and then increases again for increasing dissipation strength.

\begin{figure}[t!]
\includegraphics[width=\linewidth]{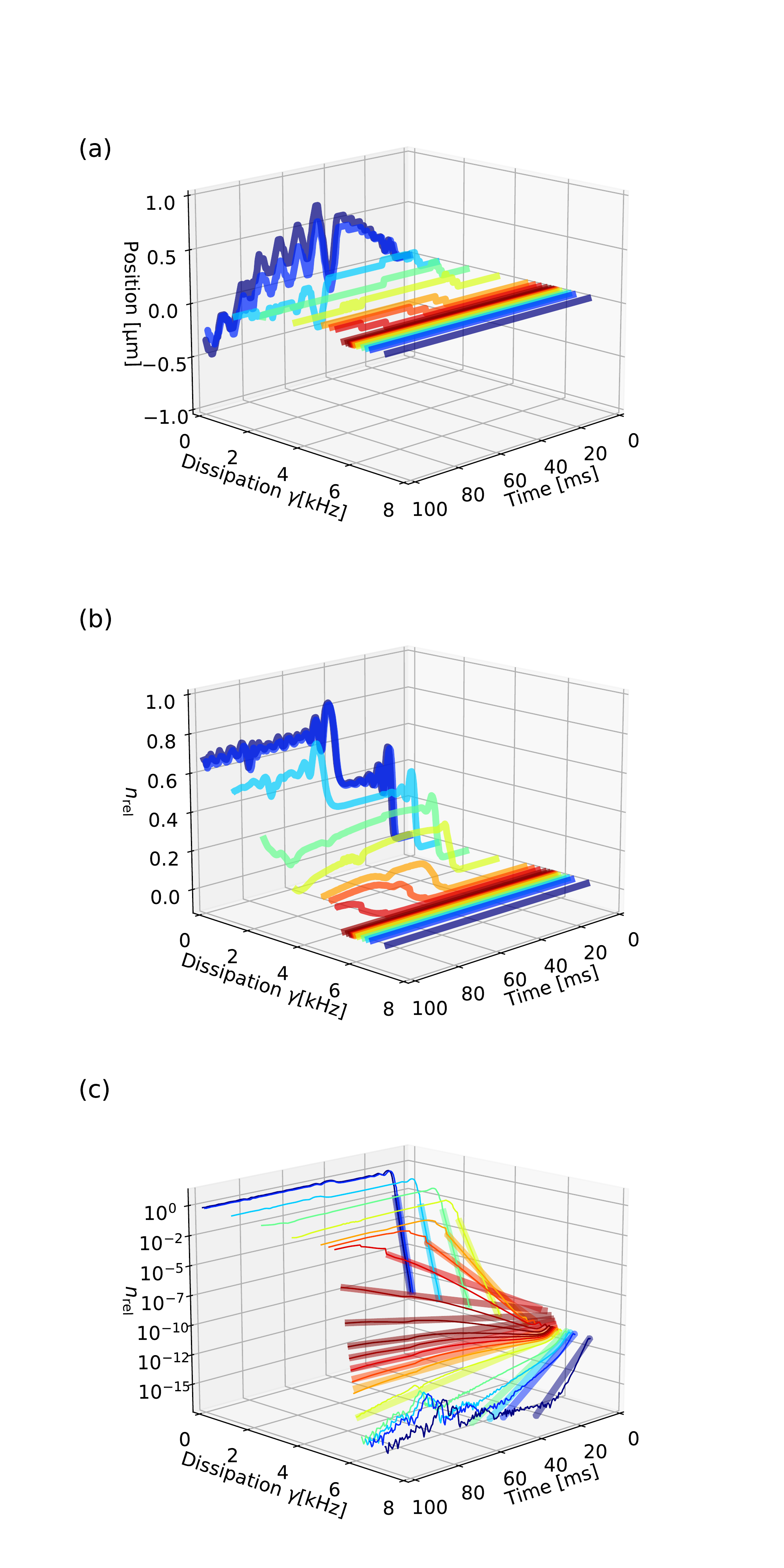}
\caption{Analysis of the decay of the DS.
(a) Time evolution of the position of the density minimum.
For large $\gamma$, the dark soliton is stable and the density minimum does not move.
For small $\gamma$, the SV shows oscillatory behavior, while for intermediate values of $\gamma$, the VR moves steadily in one direction.
(b) Evolution of the radially integrated density in the density minimum.
The three solitary waves have distinct values of the density.
(c) Exponential fit of the initial part of the dynamics to extract the decay time of the DS.}
\label{fig:densityAndPositionOfMin_gamma_t}
\end{figure}

\begin{figure}[t!]
\includegraphics[width=\linewidth]{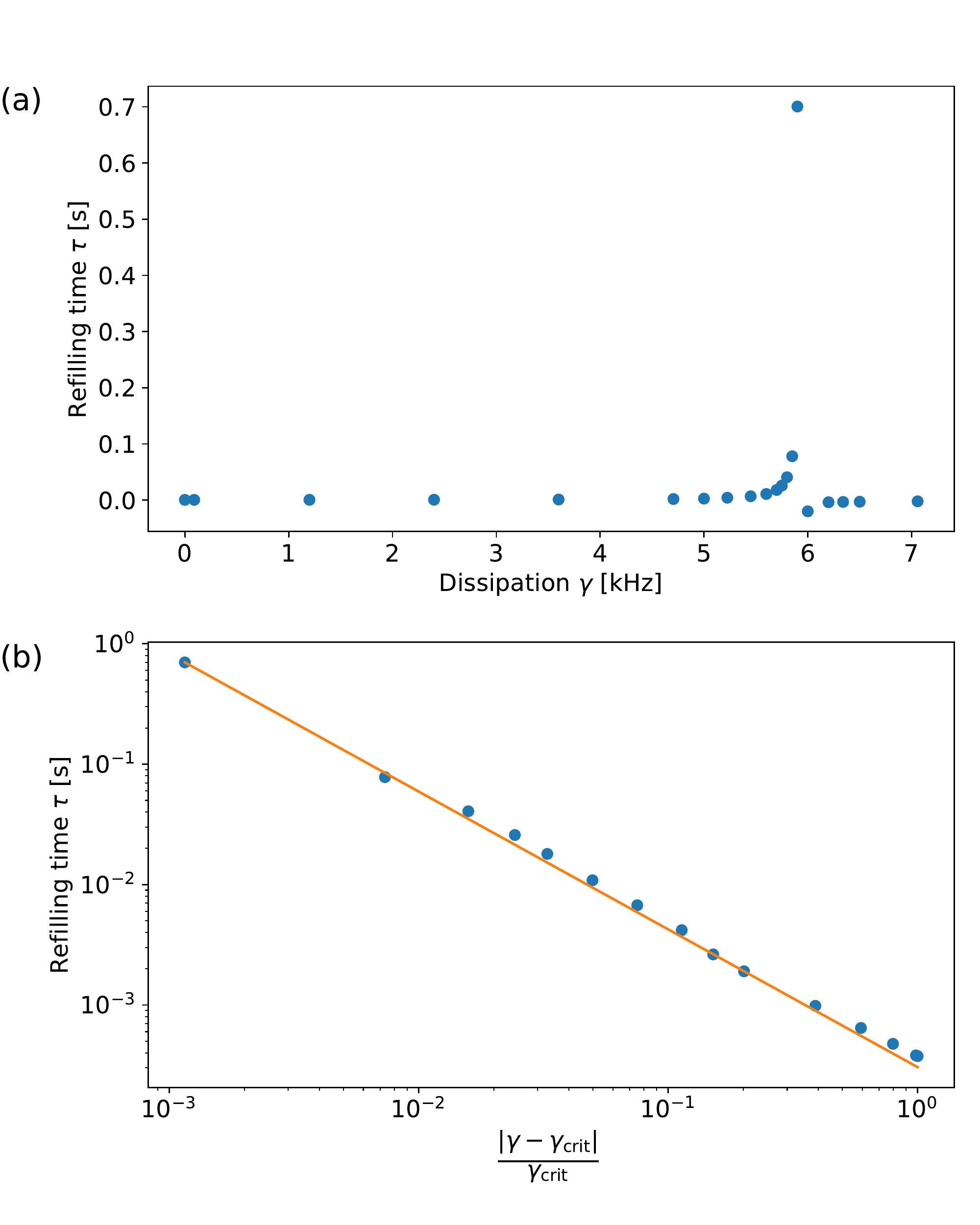}
\caption{(a) Refilling time extracted from an exponential fit to the initial dynamics of $n_{\mathrm{rel}}$ over dissipation $\gamma$ (compare Fig.\,\ref{fig:densityAndPositionOfMin_gamma_t}c).
(b) Same data rescaled and plotted logarithmically together with a power law fit.}
\label{fig:refillingTime_gamma_gammaCrit}
\end{figure}

To quantify the decay process of the DS and to determine the threshold value of $\gamma_0$ for its stabilization, we analyze the minimum density in the center of the respective steady-state.
To this end, we integrate the density along the $z$-direction, $\left\vert\tilde{\psi}\left(x,y\right)\right\vert^{2} = \sum_z \left\vert\psi\left(x,y,z\right)\right\vert^{2}$ and numerically determine the axial position of the density minimum, $x_{\mathrm{min}}=\min\left(\left\vert\tilde{\psi}\left(x,0\right)\right\vert^{2}\right)$.
The result is shown in fig. \ref{fig:densityAndPositionOfMin_gamma_t} (a).
We observe that the DS remains at its position at the center of the BEC, the VR continuously moves at very slow velocities in one direction and the SV shows oscillatory dynamics.
To further quantify the dynamics of the decay of the DS, we consider the integrated density of the slice at $x_{\mathrm{min}}$, i.e. $n = \sum_{y,z}\left\vert\psi\left(x_{\mathrm{min}},y,z\right)\right\vert^{2}$ and compare it to the integrated density of such a slice at the axial center of the ground state wave function, i.e. $n_{0} = \sum_{y,z}\left\vert\psi_{\mathrm{gs}}\left(0,y,z\right)\right\vert^{2}$.
The relative density at the position of the soliton is then defined by $n_{\mathrm{rel}} = \frac{n}{n_{0}}$.
This is shown in fig. \ref{fig:densityAndPositionOfMin_gamma_t} (b).
Again, we can identify the three different solitary waves: the DS with the lowest density, the VR with an intermediate density and the SV with the highest density.
For a DS we find $n_{\mathrm{rel}} = 2.4\cdot10^{-12}$.
To extract the decay time $\tau$ of the DS we fit the initial dynamics of $n_{\mathrm{rel}}(t)$ with an exponential function $A\cdot\exp\left(\frac{t}{\tau}\right)$.
We restrict the time to $t_{\mathrm{max}}$ where $n_{\mathrm{rel}}\left(t_{\mathrm{max}}\right) = 1\%$.
The result of the fit is shown as a solid line in fig. \ref{fig:densityAndPositionOfMin_gamma_t} (c).
One can clearly see that the initial dynamics shows an exponential behavior.
Note the large dynamic range over which the density is changing.
The initial decrease of $n_{\mathrm{rel}}$ for high $\gamma_{0}$ where the DS is stabilized in fig. \ref{fig:densityAndPositionOfMin_gamma_t} (c) originates from the fact that the DS and the imaginary potential do not perfectly overlap.
Thus there is an initial reduction of the density until the system has reached its steady state.
The fitted decay time $\tau$ is shown in fig. \ref{fig:refillingTime_gamma_gammaCrit} over $\gamma_{0}$.

We see that $\tau$ increases with increasing $\gamma_{0}$ up to a critical point, $\gamma_{\mathrm{crit}}$, before it drops.
This is typical for a critical slowing down, described by an algebraic behavior of the form 

\begin{equation}
\tau = a\left(\frac{\left\vert\gamma-\gamma_{\mathrm{crit}}\right\vert}{\gamma_{\mathrm{crit}}}\right)^{b}
\end{equation}

Fitting this model to our data, we find $\gamma_{\mathrm{crit}} = \SI{5893}{\Hz}$, $a = 3.60\cdot 10^{-4}$\,s, and a critical exponent of $b=-1.12$.  
The data and the fit are shown in Fig.\,\ref{fig:refillingTime_gamma_gammaCrit}b.

\section{Attractor dynamics towards the dark kink solitons} \label{sec:attractor}

Having established that the dark soliton is the steady-state of the system for $\gamma>\gamma_{\mathrm{crit}}$, we now analyze its attractor dynamics.
To this end, we consider two different dissipation strengths: $\gamma_{0,<} = \SI{1200}{\Hz}$ which is below the stabilization threshold for the DS and $\gamma_{0,>} = \SI{7100}{\Hz}$ which is above the stabilization threshold.
As initial conditions, we chose different grey kink solitons (GKS) and study their time evolution under the influence of dissipation.
To construct proper initial states, we start from the wave function of the ground state $\psi_{\mathrm{gs}}$ in the harmonic trap.
To obtain a wave function which is close to the GKS we multiply the ground state wave function by the function which describes a moving DS in a homogenios background, i.e.
\begin{align}
\psi_{\mathrm{GKS}}&\left(x,y,z\right) = \psi_{\mathrm{gs}}\left(x,y,z\right)\\
\nonumber
&\times\left(\tilde{v} - i\sqrt{1-\tilde{v}^{2}}\cdot\tanh\left(\frac{x}{\xi\left(x,y,z\right)}\sqrt{1-\tilde{v}^{2}}\right)\right)
\end{align}
where $\tilde{v}$ is the velocity of the GKS in units of the speed of sound.
This is related to the phase difference far away from the kink by
\begin{equation}
\Delta\phi = \phi\left(x\to\infty\right) - \phi\left(x\to-\infty\right) = -2\arccos\left(\tilde{v}\right)~.
\end{equation}

\begin{figure}[t!]
\includegraphics[width=\linewidth]{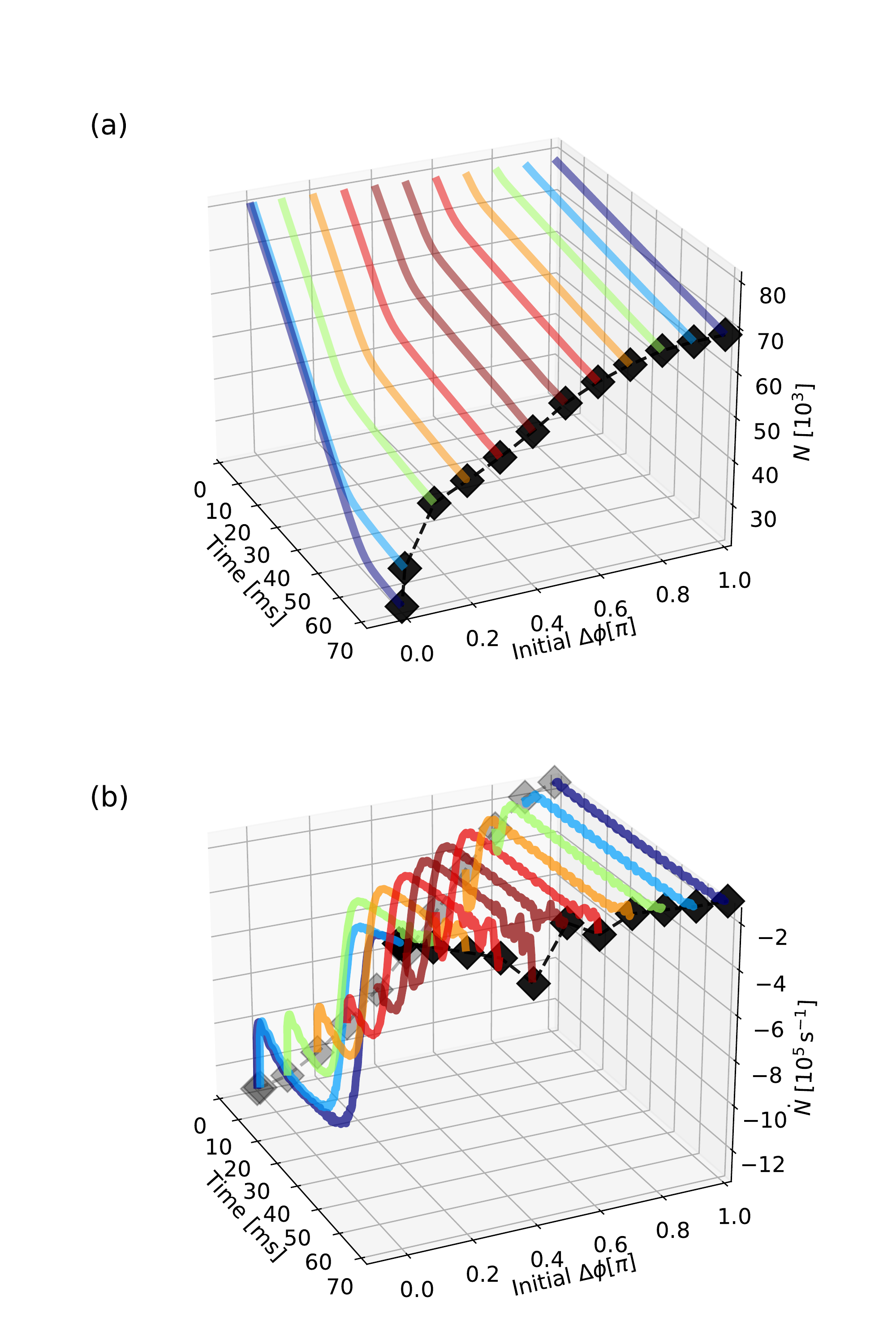}
\caption{Attractor dynamics towards the dark soliton.
(a) Number of particles $N(t)$ and (b) loss rate $\dot{N}(t)$ for different initial phase difference $\Delta\phi$ of the grey kink soliton.}
\label{fig:N_dNdt_deltaPhi_t_varPhi}
\end{figure}

\begin{figure}[t!]
\includegraphics[width=\linewidth]{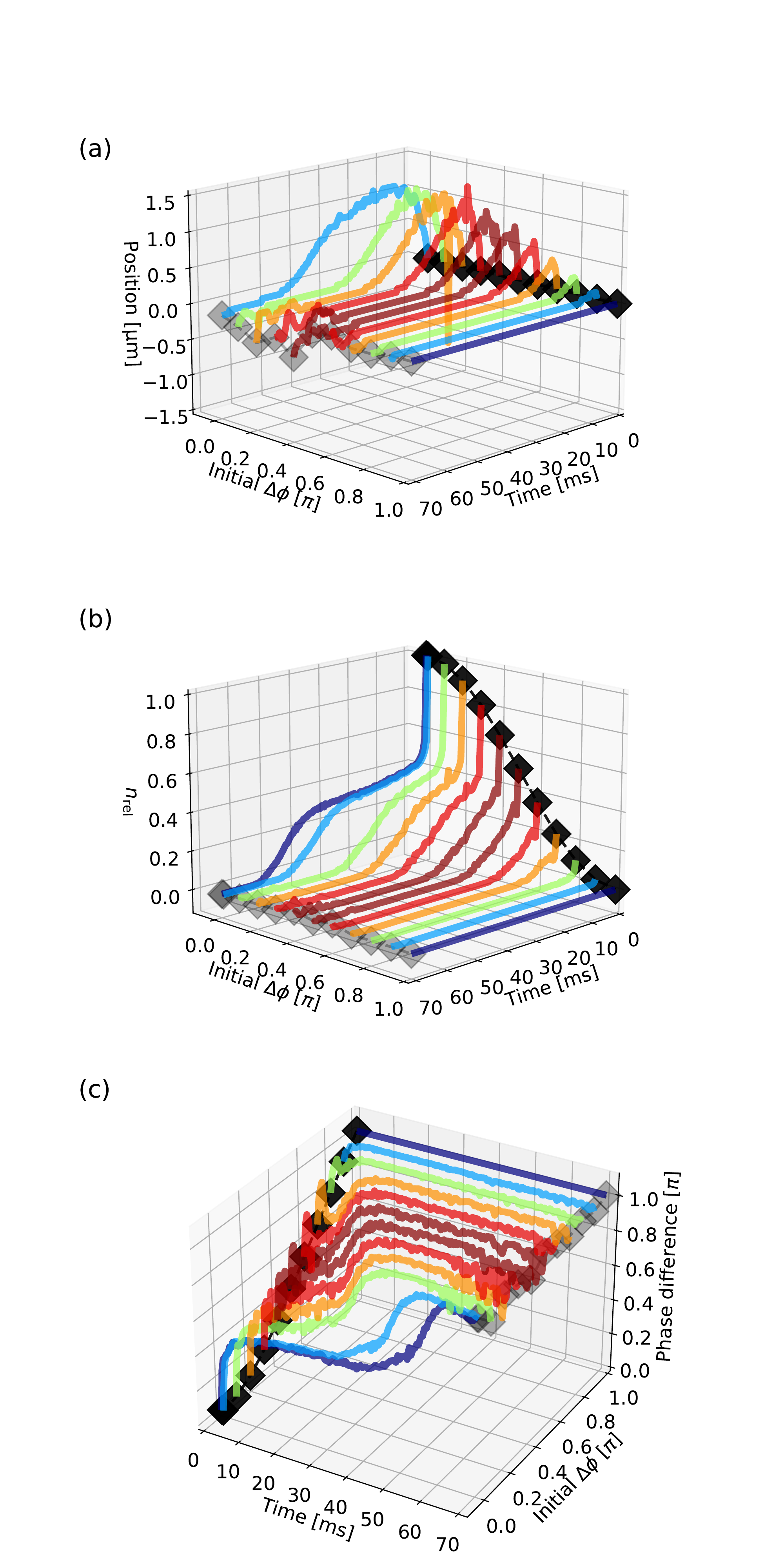}
\caption{Attractor dynamics towards the dark soliton.
(a) Evolution of the position of the density minimum.
The initial motion of the grey soliton is damped and the DS is stabilized at the central position, where the dissipation is located.
(b) The density in the minimum approaches zero.
(c) The phase difference across the density minimum approaches the characteristic value of $\pi$ of the DS.}
\label{fig:densityAndPositionOfMin_deltaPhi_t_varPhi}
\end{figure}

The local healing length is denoted by $\xi\left(x,y,z\right)$. 
In the following, we choose the full range of possible phase differences between the two ends of the wave function, $\Delta\phi\in\left[0,\pi\right]$. 
This way, we can sample all initial states interpolating between the BEC ground state and the DS.
In order to be sensitive to instability modes, we add for the evolution of the ground state 5\% of Gaussian noise before evolving in time.

For $\gamma_{0,<}$ we see qualitatively the same behavior as described in ref. \cite{Mateo_2015}: The GKS oscillates in the trap and eventually decays.
No attractor dynamics is visible.
The situation changes for $\gamma_{0,>}$.
In Fig.\,\ref{fig:N_dNdt_deltaPhi_t_varPhi} the number of particles $N(t)$ and the loss rate $\dot{N}(t)$ over time are shown for the different initial phase differences $\Delta\phi$.
The evolution of the atom number shows that the initial losses are reduced, if the GKS gets closer to the DS, which features the lowest amount of losses.
We can also see that the total atom number shows a kink, after which the losses are reduced.
This shows the appearance of a stable DS, which becomes better visible if we look at the loss rate $\dot{N}(t)$ in Fig.\,\ref{fig:N_dNdt_deltaPhi_t_varPhi}b.
For all initial $\Delta\phi$ we observe the same $\dot{N}$ after \SI{50}{\milli\second}, signaling the attraction to the same steady-state.

To further verify the attraction towards a DS we find the plane of minimal density in axial direction as described in sec. \ref{sec:stabilityDS} and plot its position (fig. \ref{fig:densityAndPositionOfMin_deltaPhi_t_varPhi} (a)) and its relative density (fig. \ref{fig:densityAndPositionOfMin_deltaPhi_t_varPhi} (b)).
We see that for low phase differences the GKS moves away from the center and is attracted back towards the location of the dissipation at $x=0$.
Also $n_{\mathrm{rel}}$ approaches the minimum value of the DS.
Finally we consider the phase difference which we define as $\Delta\phi\left(t\right) = \mathrm{arg}\left(\psi\left(x_{\mathrm{min}}+\SI{1}{\micro\meter},0,0\right)\right) - \mathrm{arg}\left(\psi\left(x_{\mathrm{min}}-\SI{1}{\micro\meter},0,0\right)\right)$.
In fig. \ref{fig:densityAndPositionOfMin_deltaPhi_t_varPhi} (c) we see that this approaches $\pi$ for all $\Delta\phi\left(t=0\right)$.
I.e. the wave function is attracted towards the DS.
Our results show that the DS is the unique steady-state for a whole class of initial states and even the condensate ground state is attracted towards it.

\section{Discussion and conclusions} \label{sec:concl}

We studied the dynamics of a 3D Bose-Einstein condensate with a dark or a grey kink soliton in an elongated trap with localized dissipation.
In the case of the dark soliton, we find that the snaking instability is suppressed above a certain threshold value of the dissipation strength.
For the grey soliton, we observe an attraction of the system towards the dark soliton.
We find that the dark soliton is the unique steady-state for all initial grey solitons, even when starting from the BEC ground state.
Performing numerical experiments, we can however not exclude, that another steady-state exists for the given parameters.
The existence of such a state would be intriguing, as one could observe and study bistable behavior in the system.
To generalize our work, it would be interesting to perform a linear stability analysis on the two situations presented here within the framework of a Bogoliubov transformation and see how the imaginary frequencies become suppressed with increasing dissipation strength.
This would help to establish the full phase diagram of the system and to relate our findings to dissipative phase transitions.

\section{Acknowledgements}
We gratefully acknowledge discussions with Joachim Brand, Antonio Mu\~noz Mateo, Corinna Kollath and Ian Spielman.
We acknowledge financial support by the DFG within the collaborative research center OSCAR, project B3 (number 277625399), and within the graduate school of excellence MAINZ.

\bibliography{soliton_stability_dissipation}

\end{document}